\documentclass[aps,12pt,prb,showpacs]{revtex4}
\usepackage{amssymb}
\usepackage{color}
\usepackage{graphicx}
\usepackage{dcolumn}
\usepackage{bm}

\begin{document}

\title{Macroscopic Quantum Tunneling Effect of $Z_{2}$ Topological Order}
\author{Jing Yu}
\affiliation{Department of Physics, Beijing Normal University, Beijing 100875, China}
\author{Su-Peng Kou}
\thanks{Corresponding author}
\email{spkou@bnu.edu.cn}
\affiliation{Department of Physics, Beijing Normal University, Beijing 100875, China}

\begin{abstract}
In this paper, macroscopic quantum tunneling (MQT) effect of $Z_{2}$
topological order in the Wen-Plaquette model is studied. This kind of MQT is
characterized by quantum tunneling processes of different virtual
quasi-particles moving around a torus. By a high-order degenerate
perturbation approach, the effective pseudo-spin models of the degenerate
ground states are obtained. From these models, we get the energy splitting
of the ground states, of which the results are consistent with those from
exact diagonalization method.
\end{abstract}

\pacs{75.45.+j, 03.67.Lx, 03.65.Xp, 75.10.Jm} \maketitle

\section{Introduction}

In quantum mechanics, quantum tunneling effect is a process by which quantum
particles penetrate barriers, which are forbidden in classical processes
\cite{Razag}. It is Gamov who pointed out that a single $\alpha $ particle
can tunnel through a barrier which introduced "macroscopic quantum
tunneling" (MQT) into physics firstly. Macroscopic quantum tunneling effects
have been widely applied to different research fields, such as quantum
oscillations between two degenerate wells of \textrm{NH}$_{3}$, quantum
coherence in one dimension charge density waves, macroscopic quantum
tunneling effect in ferromagnetic single domain magnets and quantum
tunneling phenomena in biased Josephson junctions. In general, to find MQT
in a system, there must exist two or more separated "classical" states with
macroscopically distinct. As shown in Fig.1, a quantum particle may take a
short cut from one well to the other without climbing the barrier.

\begin{figure}[tbp]
\includegraphics[clip,width=0.55\textwidth]{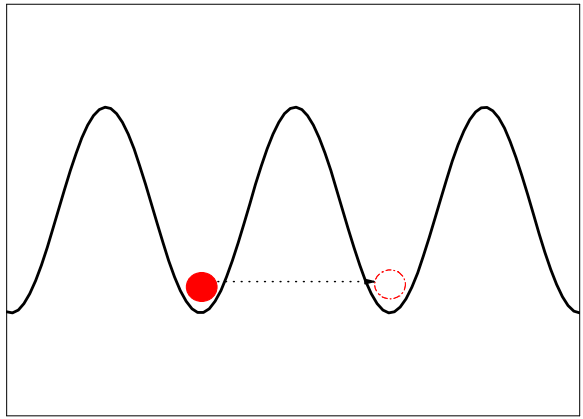}
\caption{The scheme of a typical macroscopic quantum tunneling process. }
\label{Fig.1}
\end{figure}

In this paper we will study a new class of MQT - the MQT in $Z_{2}$
topological order. At the beginning we give a brief introduction to $Z_{2}$
topological order. Topological order is a new type of quantum orders beyond
Landau's symmetry breaking paradigm\cite{wen,wen1,wen3,wen4,wen5}, of which
there are four universal properties : 1) All excitations have mass gap; 2)
The quantum degeneracy of the ground states depends on the genius of the
manifold of the background; 3) There are (closed) string net condensations;
4) Quasi-particles have exotic statistics. All these properties are robust
against perturbations. $Z_{2}$ topological order is the simplest topological
ordered state with three types of quasiparticles: $Z_{2}$ charge, $Z_{2}$
vortex, and fermions\cite{wen4}. $Z_{2}$ charge and $Z_{2}$ vortex are all
bosons with mutual $\pi $ statistics between them. The fermions can be
regarded as bound states of a $Z_{2}$ charge and a $Z_{2}$ vortex. In last
ten years, several exactly solvable spin models with $Z_{2}$ topological
orders were found, such as the Kitaev toric-code model \cite{k1}, the
Wen-plaquette model\cite{wen4,wen5} and the Kitaev model on honeycomb lattice%
\cite{k2}.

A decade ago, Kitaev pointed out that the degenerate ground states of a $%
Z_{2}$ topological order make up a protected code subspace (the so-called
toric-code) free from error\cite{k1,k2}. In Ref.\cite{ioffe}, topological
qubit based on the degenerate ground states of a $Z_{2}$ topological order
has been designed. Then one can manipulate the degenerate ground states by
braiding anyons, which has becomes a hot issue recently \cite%
{pa,sarma,du1,zoller,vid,vids,zhang,zhang1,gao,kou1,kou1',vid1,vid2}.
Recently, an alternative approach to design TQC is proposed by manipulating
the protected code subspace\cite{kou1,kou1'}. The key point to manipulate
the degenerate ground states is to tune their MQT effect. Thus it becomes an
interesting issue to study the MQT in $Z_{2}$ topological order.

In this paper, by using a high-order degenerate perturbative approach, we
study the MQT of the degenerate ground states of $Z_{2}$ topological order,
taking the Wen-plaquette Model as an example. The remainder of the paper is
organized as follows. In Sec. II, the degenerate ground states of the
Wen-plaquette model is classified by the topological closed string
operators. In Sec.III, the dynamics of quasi-particles are studied. In Sec.
IV, the MQT of the degenerate ground states of the $Z_{2}$ topological order
are formalized on a torus of different lattices. The numerical results are
given to compare with the theoretical results. Finally, the conclusions are
given in Sec. V.

\section{The degenerate ground states and its representation of string
operators}

In this section, we study the degenerate ground states of the Wen-plaquette
model. The Hamiltonian of the Wen-plaquette model is given by
\begin{equation}
\hat{H}=-g\sum_{i}\hat{F}_{i},
\end{equation}%
with
\begin{equation}
\hat{F}_{i}=\sigma _{i}^{x}\sigma _{i+\hat{e}_{x}}^{y}\sigma _{i+\hat{e}_{x}+%
\hat{e}_{y}}^{x}\sigma _{i+\hat{e}_{y}}^{y}
\end{equation}%
and $g>0.$ $\sigma _{i}^{x},$ $\sigma _{i}^{y}$ are Pauli matrices on site $%
i.$ The ground states of the Wen-plaquette model denoted by ${F_{i}\equiv +1}
$ at each site are known to be an example of $Z_{2}$ topological state. The
ground state energy becomes $E_{0}=-gN$ where $N$ is the total lattice number%
\cite{wen,wen4,wen5,kou2}.

In the topological order of the Wen-plaquette model, there exist three types
of open string operators $W_{c}(C)$, $W_{v}(C)$, $W_{f}(C)$ corresponding to
three types of quasi-particles: $Z_{2}$ charge, $Z_{2}$ vortex, and fermion,
respectively\cite{wen1}. Here $C$ is a (closed or open) loop. To create a $%
Z_{2}$ vortex (charge) excitation, one may draw a string state that connects
nearest neighboring \emph{even} (\emph{odd}) plaquettes $W_{v}(C)$ (or $%
W_{c}(C)$). Such a string state is created by the following string operator $%
\prod_{C}\sigma _{i}^{a_{i}}$ where the product $\prod_{C}$ is over all the
sites on the string along a loop $C$ connecting even-plaquettes (or
odd-plaquettes), $a_{i}=y$ if $i$ is even and $a_{i}=x$ if $i$ is odd. For a
fermionic excitation, the string operator has a form as $W_{f}(C)=\prod_{n}%
\sigma _{i_{n}}^{l_{n}}$ with a string $C$ connecting the mid-points of the
neighboring links, and $i_{n}$ are sites on the string. $l_{m}=z$ if the
string does not turn at site $i_{m}$. $l_{m}=x$ or $y$ if the string makes a
turn at site $i_{m}$. $l_{m}=y$ if the turn forms a upper-right or
lower-left corner. $l_{m}=x$ if the turn forms a lower-right or upper-left
corner. It is obvious that the fermionic string can be regarded as a bound
state of strings of the $Z_{2}$ charges and the $Z_{2}$ vortices, that is $%
W_{f}(C)=W_{c}(C)W_{v}(C)$. If $C$ are closed loops, we get condensed
closed-string operators of the ground states $|\Psi _{0}\rangle $ as
\begin{eqnarray}
\langle \Psi _{0}|W_{c}(C)|\Psi _{0}\rangle &=&1,\text{ }\langle \Psi
_{0}|W_{v}(C)|\Psi _{0}\rangle =1, \\
\langle \Psi _{0}|W_{f}(C)|\Psi _{0}\rangle &=&1.  \nonumber
\end{eqnarray}%
One can see the detailed definition of the string operators in Ref.\cite%
{wen1}.

\begin{figure}[tbp]
\includegraphics[clip,width=0.6\textwidth]{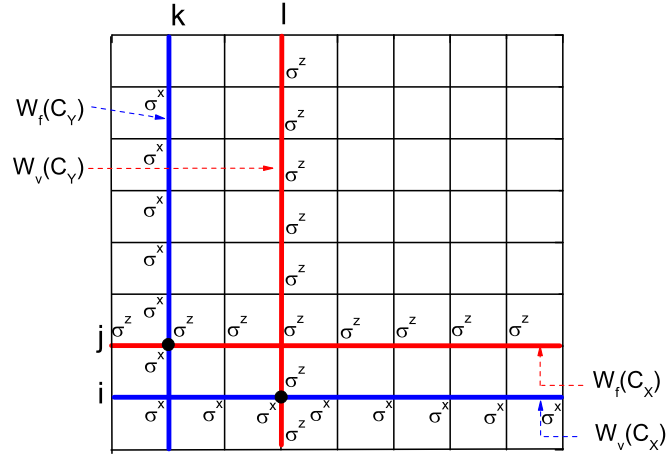}
\caption{The topological closed string operators on a torus. The dots denote
the crosses of different types of strings.}
\label{Fig.2}
\end{figure}

To classify the degeneracy of the ground states, we define three types of
topological closed-string operators $W_{c}(\mathcal{C}),$ $W_{v}(\mathcal{C}%
) $ and $W_{f}(\mathcal{C})=W_{c}(\mathcal{C})W_{v}(\mathcal{C})$, with $%
\mathcal{C}$ denoting topological closed loops. The word $^{\prime }$\emph{%
topological}$^{\prime }$ means that the $^{\prime }$\emph{big}$^{\prime }$
loops $\mathcal{C}$ surround the torus globally (See Fig.2). One can easily
check the commutation relations between the topological closed string
operators and the Hamiltonian
\begin{equation}
\left[ H,\text{ }W_{c}(\mathcal{C})\right] =\left[ H,\text{ }W_{v}(\mathcal{C%
})\right] =\left[ H,\text{ }W_{f}(\mathcal{C})\right] =0.
\end{equation}

{F}or the ground states on a torus of an even-by-even ($e\ast e$) lattice,
we can define four types of elementary topological closed string operators, $%
W_{v}(\mathcal{C}_{X}),$ $W_{v}(\mathcal{C}_{Y}),$ $W_{f}(\mathcal{C}_{X})$
and $W_{f}(\mathcal{C}_{Y}).$ Here $\mathcal{C}_{X}$ denotes a closed-loop
around the torus along $e_{x}$-direction and $\mathcal{C}_{Y}$ denotes a
closed loop around the torus along $e_{y}$-direction. Due to the commutation
(or anti-commutation) relations between them
\begin{eqnarray}
\left[ W_{v}(\mathcal{C}_{X}),W_{f}(\mathcal{C}_{X})\right] &=&0,\text{ }%
\left[ W_{v}(\mathcal{C}_{Y}),W_{f}(\mathcal{C}_{Y})\right] =0, \\
\left[ W_{v}(\mathcal{C}_{X}),W_{v}(\mathcal{C}_{Y})\right] &=&0,\text{ }%
\left[ W_{f}(\mathcal{C}_{X}),W_{f}(\mathcal{C}_{Y})\right] =0,  \nonumber \\
\left \{ W_{v}(\mathcal{C}_{X}),W_{f}(\mathcal{C}_{Y})\right \} &=&0,\text{ }%
\left \{ W_{v}(\mathcal{C}_{Y}),W_{f}(\mathcal{C}_{X})\right \} =0,
\nonumber
\end{eqnarray}%
\ we may identify $W_{v}(\mathcal{C}_{X}),$ $W_{v}(\mathcal{C}_{Y}),$ $W_{f}(%
\mathcal{C}_{X})$ and $W_{f}(\mathcal{C}_{Y})$ by pseudo-spin operators $%
\tau _{1}^{x},$ $\tau _{2}^{x}$ $\tau _{2}^{z},$ and $\tau _{1}^{z}$ as
\begin{eqnarray}
W_{v}(\mathcal{C}_{X}) &=&\tau _{1}^{x}\otimes \mathbf{1},\text{ }W_{v}(%
\mathcal{C}_{Y})\rightarrow \mathbf{1}\otimes \tau _{2}^{x}, \\
W_{f}(\mathcal{C}_{X}) &\rightarrow &\mathbf{1}\otimes \tau _{2}^{z},\text{ }%
W_{f}(\mathcal{C}_{Y})\rightarrow \tau _{1}^{z}\otimes \mathbf{1}.  \nonumber
\end{eqnarray}%
Thus other five topological closed string operators $W_{c}(\mathcal{C}_{X}),$
$W_{c}(\mathcal{C}_{Y}),$ $W_{c}(\mathcal{C}_{XY}),$ $W_{v}(\mathcal{C}%
_{XY}) $ and $W_{f}(\mathcal{C}_{XY})$ are denoted by $\tau _{1}^{x}\otimes
\tau _{2}^{z},$ $\tau _{1}^{z}\otimes \tau _{2}^{x},$ $\tau _{1}^{y}\otimes
\tau _{2}^{y},$ $\tau _{1}^{x}\otimes \tau _{2}^{x}$ and $\tau
_{1}^{z}\otimes \tau _{2}^{z}$, respectively,%
\begin{eqnarray}
W_{c}(\mathcal{C}_{X}) &\rightarrow &\tau _{1}^{x}\otimes \tau _{2}^{z},%
\text{ }W_{c}(\mathcal{C}_{Y})\rightarrow \tau _{1}^{z}\otimes \tau _{2}^{x},
\\
W_{c}(\mathcal{C}_{XY}) &\rightarrow &\tau _{1}^{y}\otimes \tau _{2}^{y},%
\text{ }W_{v}(\mathcal{C}_{XY})\rightarrow \tau _{1}^{x}\otimes \tau
_{2}^{x},  \nonumber \\
W_{f}(\mathcal{C}_{XY}) &\rightarrow &\tau _{1}^{z}\otimes \tau _{2}^{z}.
\nonumber
\end{eqnarray}%
Here $\mathcal{C}_{XY}$ is a closed loop around the torus along diagonal
directions. In the table.(I), the pseudo-spin representation of the
topological closed string operators are illustrated.

\begin{table}[t]
\begin{tabular}{|c|cccc|}
\hline
Pesudo-spin operators & $\mathcal{C}_{X}$ & $\mathcal{C}_{Y}$ & $\mathcal{C}%
_{XY}$ &  \\ \hline
$Z_{2}$-vortex & $\tau _{1}^{x}\otimes \mathbf{1}$ & $\mathbf{1}\otimes \tau
_{2}^{x}$ & $\tau _{1}^{x}\otimes \tau _{2}^{x}$ &  \\
$Z_{2}$-charge & $\tau _{1}^{x}\otimes \tau _{2}^{z}$ & $\tau
_{1}^{z}\otimes \tau _{2}^{x}$ & $\tau _{1}^{y}\otimes \tau _{2}^{y}$ &  \\
$Fermion$ & $\mathbf{1}\otimes \tau _{2}^{z}$ & $\tau _{1}^{z}\otimes
\mathbf{1}$ & $\tau _{1}^{z}\otimes \tau _{2}^{z}$ &  \\ \hline
\end{tabular}%
\caption{Pseudo-spin representation of the topological closed string
operators on an even-by-even lattice.}
\label{ee}
\end{table}

Then as the eigenstates of $\tau _{l}^{z}$ ($l=1,2$), the four degenerate
ground states are denoted by $\mid m_{1},m_{2}\rangle =\mid m_{1}\rangle
\otimes \mid m_{2}\rangle $. For $m_{l}=0,$ we have
\begin{equation}
\tau _{l}^{z}\mid m_{l}\rangle =\mid m_{l}\rangle ,
\end{equation}%
and for $m_{l}=1$ we have
\begin{equation}
\tau _{l}^{z}\mid m_{l}\rangle =-\mid m_{l}\rangle .
\end{equation}%
Physically, the topological degeneracy arises from presence or the absence
of $\pi $ flux of fermion through the hole. The values of $m_{l}$ reflect
the presence ($m_{l}=1$) or the absence ($m_{l}=0$) of the $\pi $ flux in
the hole.

For the degenerate ground states on an even-by-odd ($e\ast o$) lattice, the
situation changes. Because a $Z_{2}$ vortex or $Z_{2}$ charge has to move
even steps to go back to the same plaquette around a torus, we cannot well
define a topological closed string operator of $Z_{2}$ vortex or $Z_{2}$
charge along $e_{y}$-direction, of which the loop consists of odd number
plaquettes. So we can only define topological closed string operator of $%
Z_{2}$ vortex and $Z_{2}$ charge along $e_{x}$-direction $W(\mathcal{C}_{X})$
($W(\mathcal{C}_{X})=W_{v}(\mathcal{C}_{X})=W_{c}(\mathcal{C}_{X})$) and the
corresponding fermionic string operator along $e_{y}$-direction $W_{f}(%
\mathcal{C}_{Y}).$ Due to the anti-commutation relations between $W(\mathcal{%
C}_{X})$ and $W_{f}(\mathcal{C}_{Y})$,
\begin{equation}
\left \{ W(\mathcal{C}_{X}),\text{ }W_{f}(\mathcal{C}_{Y})\right \} =0,
\end{equation}%
\ we may represent $W(\mathcal{C}_{X})$ and $W_{f}(\mathcal{C}_{Y})$ by
pseudo-spin operators $\tau _{1}^{x}$ and $\tau _{1}^{z}$, respectively%
\begin{equation}
W(\mathcal{C}_{X})\rightarrow \tau _{1}^{x},\text{ }W_{f}(\mathcal{C}%
_{Y})\rightarrow \tau _{1}^{z}.
\end{equation}%
Therefore there are two degenerate ground states $\mid m_{1}\rangle $ that
are the eigenstates of $\tau _{1}^{z}$. In the table.(II), the pseudo-spin
representation of the topological closed string operators on $e\ast o$
lattice are shown.

\begin{table}[t]
\begin{tabular}{|c|cccc|}
\hline
Pesudo-spin operators & $\mathcal{C}_{X}$ & $\mathcal{C}_{Y}$ & $\mathcal{C}%
_{XY}$ &  \\ \hline
$Z_{2}$-vortex & $\tau _{1}^{x}$ & $-$ & $\tau _{1}^{x}$ &  \\
$Z_{2}$-charge & $\tau _{1}^{x}$ & $-$ & $\tau _{1}^{x}$ &  \\
$Fermion$ & $1$ & $\tau _{1}^{z}$ & $\tau _{1}^{z}$ &  \\ \hline
\end{tabular}%
\caption{Pseudo-spin representation of the topological closed string
operators on an even-by-odd lattice.}
\label{eo}
\end{table}

Similarly, for the degenerate ground states on an odd-by-even ($o\ast e$)
lattice there are also two types of closed string operators, $W(\mathcal{C}%
_{Y})$ ($W(\mathcal{C}_{Y})=W_{v}(\mathcal{C}_{Y})=W_{c}(\mathcal{C}_{Y})$)
and $W_{f}(\mathcal{C}_{X})$, which can be described by pseudo-spin
operators $\tau _{2}^{x}$ and $\tau _{2}^{z}.$ Therefore, the two degenerate
ground states on an $o\ast e$ lattice are denoted by $\mid m_{2}\rangle $
which are the eigenstates of $\tau _{2}^{z}.$

For the degenerate ground states on an odd-by-odd ($o\ast o$) lattice, since
the total lattice number is odd, we cannot well define $Z_{2}$ vortex or $%
Z_{2}$ charge globally any more. Instead, we can only define a mixed
topological closed-string operator, $W(\mathcal{C}_{XY})=\prod_{\mathcal{C}%
}\sigma _{i}^{a_{i}}$ where the product $\prod_{\mathcal{C}}$ is over all
the sites on the string along a diagonal loop $\mathcal{C}$ connecting
plaquettes. The index $a_{i}=x$ or $y$ is determined by the position of the
plaquettes. Because $W(\mathcal{C}_{XY})$ anti-commutes with $W_{f}(\mathcal{%
C}_{X})$ and $W_{f}(\mathcal{C}_{Y})$,
\begin{equation}
\left \{ W(\mathcal{C}_{XY}),\text{ }W_{f}(\mathcal{C}_{X})\right \} =0,%
\text{ }\left \{ W(\mathcal{C}_{XY}),\text{ }W_{f}(\mathcal{C}_{Y})\right \}
=0,
\end{equation}%
we may represent $W(\mathcal{C}_{XY})$ and $W_{f}(\mathcal{C}_{X})$ (or $%
W_{f}(\mathcal{C}_{Y})$) by pseudo-spin operators $\tau ^{x}$ and $\tau ^{z}$%
, respectively%
\begin{eqnarray*}
W(\mathcal{C}_{XY}) &\rightarrow &\tau ^{x}, \\
W_{f}(\mathcal{C}_{X}) &=&W_{f}(\mathcal{C}_{Y})\rightarrow \tau ^{z}.
\end{eqnarray*}%
It is noted that
\begin{equation}
W_{f}(\mathcal{C}_{XY})=W_{f}(\mathcal{C}_{X})W_{f}(\mathcal{C}_{Y})=1.
\end{equation}%
Thus the two degenerate ground states on an $o\ast o$ lattice $\mid m\rangle
$ are the eigenstates of $\tau ^{z}$.\ In the table.(III), the pseudo-spin
representation of the topological closed string operators on $o\ast o$
lattice are shown.

\begin{table}[t]
\begin{tabular}{|c|cccc|}
\hline
Pesudo-spin operators & $\mathcal{C}_{X}$ & $\mathcal{C}_{Y}$ & $\mathcal{C}%
_{XY}$ &  \\ \hline
$Z_{2}$-vortex ($Z_{2}$-charge) & $-$ & $-$ & $\tau ^{x}$ &  \\
$Fermion$ & $\tau ^{z}$ & $\tau ^{z}$ & $1$ &  \\ \hline
\end{tabular}%
\caption{Pseudo-spin representation of the topological closed string
operators on an odd-by-odd lattice.}
\label{oo}
\end{table}

As a result, the degeneracy $\mathcal{Q}$ of the ground states of the
Wen-plaquette model on lattices with periodic boundary condition (on a
torus) is dependent on the lattice numbers : $\mathcal{Q}=4$ on $e\ast e$
lattice, $\mathcal{Q}=2$ on other cases ($e\ast o$, $o\ast e$ and $o\ast o$
lattices)\cite{wen,wen4,wen5,kou1,kou1',kou2}.

\section{Properties of quasi-particles of the Wen-plaquette model}

In this section we study the properties of the quasi-particles of the
Wen-plaquette model. In this model, $Z_{2}$ vortex is defined as ${F_{i}=-1}$
at even sub-plaquette and $Z_{2}$ charge is ${F_{i}=-1}$ at odd
sub-plaquette. The energy gap of $Z_{2}$ charge and $Z_{2}$ vortex is $2g$.
The fermions that are the bound states of a $Z_{2}$ charge and a $Z_{2}$
vortex on two neighbor plaquettes have an energy gap of $4g$. All
quasi-particles in such an exactly solvable model have flat bands. The
energy spectrums are $E_{v}=E_{c}=2g$ for $Z_{2}$ vortex and $Z_{2}$ charge,
$E_{f}=4g$ for fermions, respectively. In other words, the quasi-particles
cannot move at all. In particular, there exist two types of fermions : the
fermions on the vertical links and the fermions on the parallel links.

\begin{figure}[tbp]
\includegraphics[clip,width=0.45\textwidth]{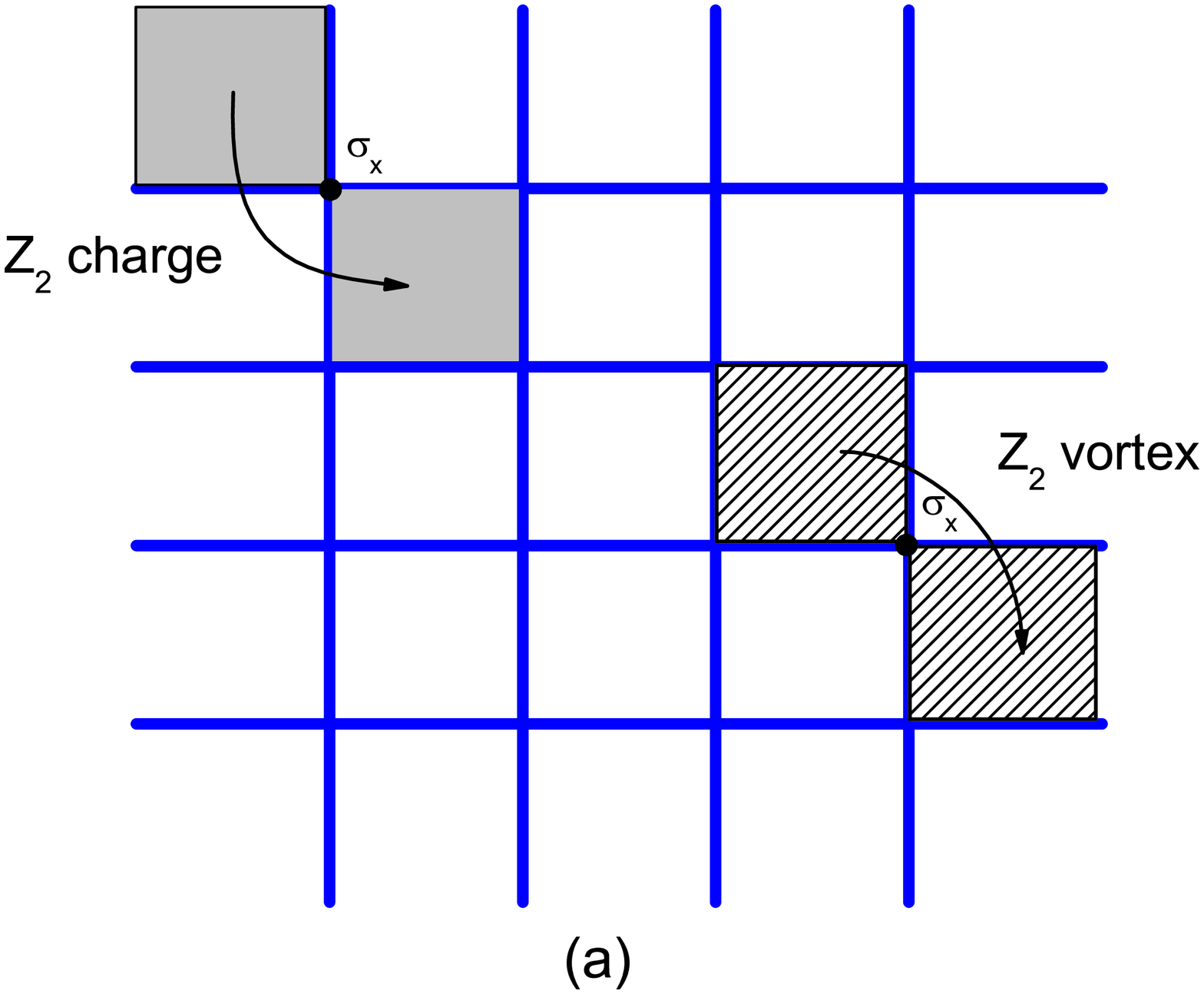} %
\includegraphics[clip,width=0.45\textwidth]{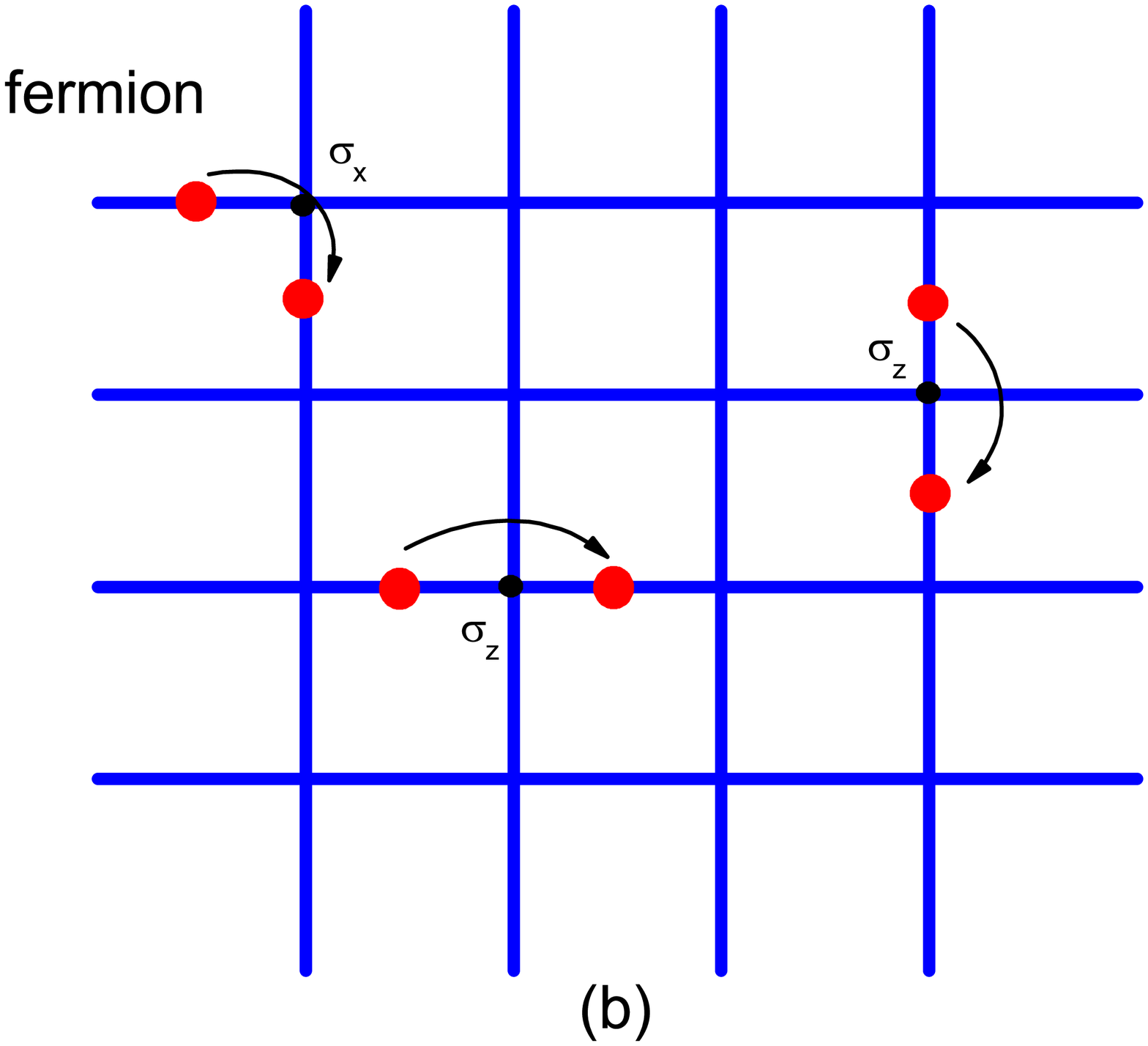}
\caption{The hoppings of $Z_{2}$ vortex, $Z2$ charge and fermions. The
shadow plaquettes, the striped plaquettes and the dots on the links
represent $Z_{2}$ vortices, $Z_{2}$ charges and fermions, respectively.}
\label{Fig.3}
\end{figure}

Under the perturbation
\begin{equation}
H_{I}=h^{x}\sum \limits_{i}\sigma _{i}^{x}+h^{z}\sum \limits_{i}\sigma
_{i}^{z},
\end{equation}%
the quasi-particles ($Z_{2}$ vortex, $Z_{2}$ charge and fermion) begin to hop%
\cite{zoller,vid,vids,vid1,vid2,kou1,kou1',kou0}. The term $h^{x}\sum
\limits_{i}\sigma _{i}^{x}$ drives the $Z_{2}$ vortex, $Z_{2}$ charge and
fermion hopping along diagonal direction $\hat{e}_{x}-\hat{e}_{y}$ (See
Fig.3(a)). For example, for a $Z_{2}$ vortex living at $i$ plaquette ${%
F_{i}=-1,}$ when $\sigma _{i}^{x}$ acts on $i+\hat{e}_{x}$ site, it hops to $%
i+\hat{e}_{x}-\hat{e}_{y}$ plaquette denoted by ${F_{i+\hat{e}_{x}-\hat{e}%
_{y}}=-1,}$%
\begin{equation}
{F_{i}}{=-1\rightarrow F_{i}=+1,}\text{ }{F_{i+\hat{e}_{x}-\hat{e}_{y}}}{%
=+1\rightarrow F_{i+\hat{e}_{x}-\hat{e}_{y}}=-1}\text{.}
\end{equation}
A pair of $Z_{2}$ vortices at $i$ and $i+\hat{e}_{x}-\hat{e}_{y}$ plaquettes
can be created or annihilated by the operation of $\sigma _{i}^{x}$,
\begin{equation}
{F_{i}}{=+1\rightarrow F_{i}=-1,}\text{ }F{_{i+\hat{e}_{x}-\hat{e}_{y}}}{%
=+1\rightarrow F_{i+\hat{e}_{x}-\hat{e}_{y}}=-1.}
\end{equation}
The term $h^{z}\sum \limits_{i}\sigma _{i}^{z}$ drive fermion hopping along $%
\hat{e}_{x}$ and $\hat{e}_{y}$ directions without affecting $Z_{2}$ vortex
and $Z_{2}$ charge : the fermions on the vertical links move along vertical
directions and the fermions on the parallel links move along parallel
directions. With the help of the term $h^{x}\sum \limits_{i}\sigma _{i}^{x},$
the two types of fermions are mixed and the fermions may turn round from
vertical links to parallel links (See Fig.3(b)).

A fact is that \emph{the topological closed string operators can be
considered as quantum tunneling processes of virtual quasi-particle moving
along the same loops}. Let us take the quantum tunneling process of $Z_{2}$
vortex as an example : at first a pair of $Z_{2}$ vortices are created. One $%
Z_{2}$ vortex propagates around the torus driven by operators $\sigma
_{i}^{x}$ and annihilates with the other $Z_{2}$ vortex. Then a string of $%
\sigma _{i}^{x}$ is left on the tunneling path, which is just the
topological closed string operator $W_{v}(\mathcal{C})$. Such a process
effectively adds a unit of a $\pi $-flux to one hole of the torus and
changes $m_{l}$ by $1$.

\section{Macroscopic quantum tunneling effects of the degenerate ground
states}

It is known that the degenerate ground states of $Z_{2}$ topological orders
have the same energy in the thermodynamic limit. The different ground states
can not mix into each other through any local fluctuations. However, in a
finite system,$\ $the degeneracy of the ground states can be (partially)
removed due to quantum tunneling processes, of which virtual quasi-particles
move around the torus\cite{k1,wen,ioffe,kou1,kou1'}. In general cases, one
will get large energy gaps for all quasi-particles and very tiny energy
splitting of the degenerate ground states $\Delta E$. Based on such
condition, we may ignore high energy excited states and consider only the
degenerate ground states. Thus in the following parts we only focus on the
ground states that are a four-level (or two-level) system.

\subsection{The high-order degenerate perturbation theory}

To solve quantum tunneling problems, people have developed many approaches
including the well known WKB (Wentzel, Kramers and Brillouin) method and the
instanton approach lately. However, based on semi-classical approximation
both above approaches are not available to the MQT of $Z_{2}$ topological
order. Instead, in this part, we develop a high-order degenerate
perturbative approach to calculate the MQT.

The Hamiltonian of the Wen-plaquette model under the external field has a
form as
\begin{equation}
\hat{H}=\hat{H_{0}}+\hat{H}_{I}
\end{equation}%
in which $\hat{H_{0}}=-g\sum_{i}\sigma _{i}^{x}\sigma _{i+\hat{e}%
_{x}}^{y}\sigma _{i+\hat{e}_{x}+\hat{e}_{y}}^{x}\sigma _{i+\hat{e}_{y}}^{y}$
is the unperturbation term, and $\hat{H}_{I}=h^{x}\sum \limits_{i}\sigma
_{i}^{x}+h^{z}\sum \limits_{i}\sigma _{i}^{z}$ is the small perturbation
one. For simplicity, we consider the quantum tunneling process between two
degenerate ground states $\mid m\rangle $ and $\mid n\rangle ,$
\begin{equation}
\mid m\rangle \Leftrightarrow \mid n\rangle .
\end{equation}

According to the Gell-Mann-Low theory, we define a transformation operator $%
\hat{U_{I}}(0,-\infty )$ as
\begin{equation}
\hat{U_{I}}(0,-\infty )=\mathrm{T}\exp (-i\int_{-\infty }^{0}\hat{H}%
_{I}^{\prime }(t^{\prime })dt^{\prime })  \label{zero}
\end{equation}%
where
\begin{equation}
\hat{H}_{I}^{\prime }(t)=e^{i\hat{H_{0}}t}\hat{H}_{I}e^{-i\hat{H_{0}}t}.
\end{equation}%
Here $\mathrm{T}$ denotes a time order and $\hslash =1$. Then the
transformation operator $\hat{U_{I}}(0,-\infty )$ in Eq.(\ref{zero}) can be
written as
\begin{equation}
\hat{U_{I}}(0,-\infty )\left \vert m\right \rangle =\sum
\limits_{j=0}^{\infty }\hat{U}_{I}^{(j)}(0,-\infty )\left \vert m\right
\rangle ,  \label{U0}
\end{equation}%
where%
\begin{eqnarray}
\hat{U}_{I}^{(0)}(0,-\infty )\left \vert m\right \rangle &=&\left \vert
m\right \rangle ,  \nonumber \\
\hat{U}_{I}^{(1)}(0,-\infty )\left \vert m\right \rangle &=&-i\int_{-\infty
}^{0}\hat{H}_{I}^{\prime }(t)dt\left \vert m\right \rangle  \nonumber \\
&=&\frac{1}{E_{0}-\hat{H_{0}}}\hat{H}_{I}\left \vert m\right \rangle ,
\nonumber \\
\hat{U}_{I}^{(2)}(0,-\infty )\left \vert m\right \rangle &=&-i\int_{-\infty
}^{0}\hat{H}_{I}^{\prime }(t)\hat{U}_{I}^{(1)}(0,-\infty )dt\left \vert
m\right \rangle  \nonumber \\
&=&\frac{1}{E_{0}-\hat{H_{0}}}\hat{H}_{I}\frac{1}{E_{0}-\hat{H_{0}}}\hat{H}%
_{I}\left \vert m\right \rangle ,  \nonumber \\
\hat{U}_{I}^{(j\neq 0)}(0,-\infty )\left \vert m\right \rangle &=&(\frac{1}{%
E_{0}-\hat{H}_{0}}\hat{H_{I}})^{j}\left \vert m\right \rangle .
\end{eqnarray}%
The element of the transformation matrix from the state $\left \vert
m\right
\rangle $ to $\left \vert n\right \rangle $ becomes
\[
\left \langle n\right \vert \hat{U_{I}}(0,-\infty )\left \vert
m\right
\rangle
\]%
and the corresponding energy is obtained as
\begin{equation}
E=\left \langle n\right \vert \hat{H}\hat{U_{I}}(0,-\infty )\left \vert
m\right \rangle =E_{0}+\delta E  \label{deltE}
\end{equation}%
where $E_{0}$ is the eigenvalue of the Hamiltonian $\hat{H}_{0}$ of $%
\left
\vert m\right \rangle .$

For the tunneling process from $\mid m\rangle $ to $\mid n\rangle ,$ a
quasi-particle will move around the torus that leads to topological closed
string operator behind. So in the sum of $j,$ the dominated term is labeled
by $j=L-1.$ $L$ is the length of the loop of a topological string operator $%
W_{\upsilon }(\mathcal{C}_{\Lambda })$ where $\upsilon =v,$ $c$ or $f$ and $%
\Lambda =X,$ $Y$ or $XY$. Then considering the tunneling process
corresponding to $W_{\upsilon }(\mathcal{C}_{\Lambda })$, we obtain the
perturbative energy as
\begin{eqnarray}
\delta E &=&\langle n\mid \hat{H_{I}}\hat{U_{I}}(0,-\infty )\mid m\rangle \\
&=&\langle n\mid \hat{H_{I}}\sum \limits_{j=0}^{\infty }\hat{U}%
_{I}^{(j)}(0,-\infty )\left \vert m\right \rangle  \nonumber \\
&=&\langle n\mid \hat{H_{I}}\hat{U}_{I}^{(L-1)}(0,-\infty )\left \vert
m\right \rangle  \nonumber
\end{eqnarray}%
Now it is noted that the operator $\hat{H_{I}}\hat{U}_{I}^{(L-1)}(0,-\infty
) $ is proportion to a topological string operator $W_{\upsilon }(\mathcal{C}%
_{\Lambda }).$

Considering all tunneling processes, we may denote the ground state energies
as a four-by-four matrix (for the four degenerate ground states on $e\ast e$
lattice) or two-by-two matrix (for the two degenerate ground states on $%
e\ast o$, $o\ast e$ and $o\ast o$ lattices),
\begin{equation}
\delta E=\sum_{m,n}\langle n\mid \hat{H_{I}}\hat{U}_{I}^{(L-1)}(0,-\infty
)\left \vert m\right \rangle .  \label{sp}
\end{equation}%
Finally we can diagonalize the four-by-four or two-by-two matrices and
obtain the energy splitting.

\subsection{Macroscopic quantum tunneling effect of the degenerate ground
states on $o\ast o$ lattice}

Firstly, we study the MQT of the two degenerate ground states on an $%
L_{x}\times L_{y}$ ($L_{x}$ and $L_{y}$ are odd numbers and $L_{x}\geq L_{y}$%
) lattice. For simplicity, we use $\mid \uparrow \rangle $ and $\mid
\downarrow \rangle $ to describe the two degenerate ground states $\mid
m=0\rangle $ and $\mid m=1\rangle ,$ respectively, of which the two ground
states can be mapped onto quantum states of pseudo-spin $\mathbf{\hat{\tau}}%
. $ Under the perturbation, $\hat{H}_{I}=h^{x}\sum \limits_{i}\sigma
_{i}^{x}+h^{z}\sum \limits_{i}\sigma _{i}^{z}$, two types of quantum
tunneling processes dominate - the one that $Z_{2}$ vortex (or $Z_{2}$
charge) propagates around the torus along diagonal direction and the other
that fermion propagates around the torus along $e_{y}$-direction.

For the first tunneling process, a virtual $Z_{2}$ vortex (or $Z_{2}$
charge) will run around the torus as long as a path with length $L_{0}$ that
is equal to $\frac{L_{x}L_{y}}{\xi }.$ Here $\xi $ is the maximum common
divisor for $L_{x}$ and $L_{y}$. For example, on a $3\times 3$ lattice, we
get $L_{0}=\frac{3\times 3}{3}=3;$ on a $3\times 5$ lattice, we get $L_{0}=%
\frac{5\times 3}{1}=15.$

\begin{figure}[tbp]
\includegraphics[clip,width=0.5\textwidth]{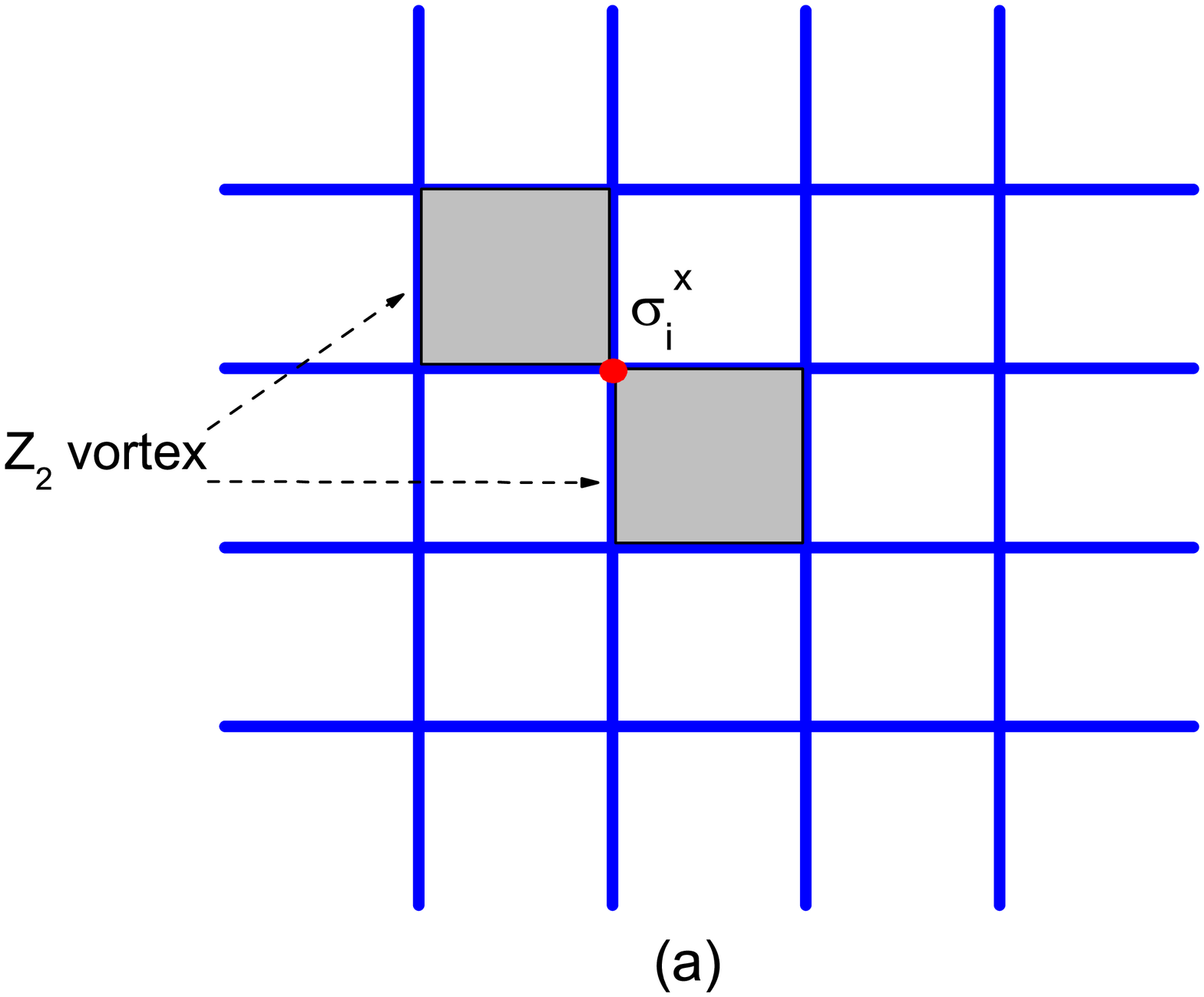} %
\includegraphics[clip,width=0.5\textwidth]{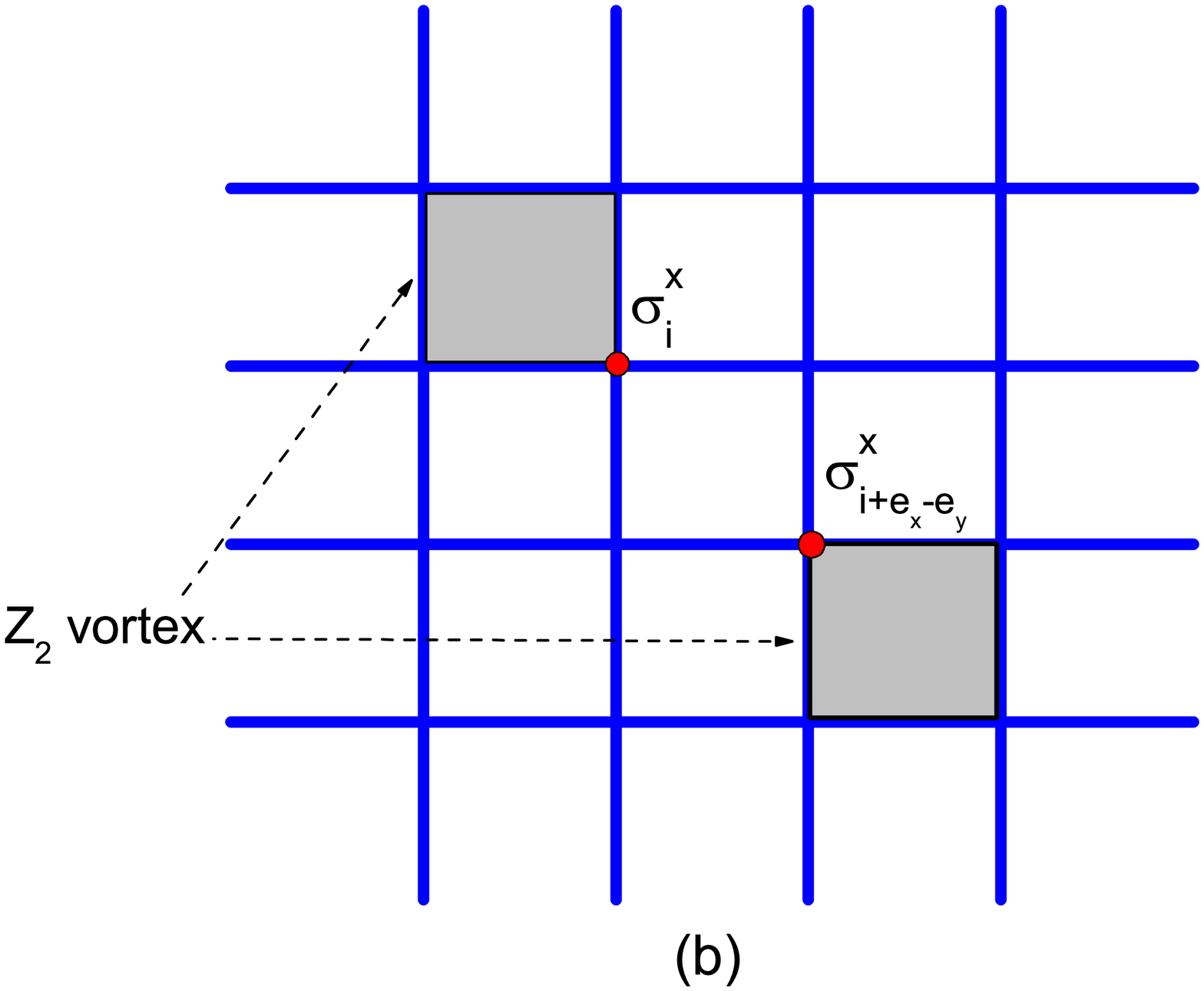}
\caption{Generation and Hopping of $Z_{2}$ vortex. The shadow plaquettes
represent $Z_{2}$ vortices.}
\label{Fig.4}
\end{figure}

From Eq.(\ref{sp}), one may obtain the energy splitting $\Delta E$ of the
two ground states as%
\begin{equation}
\delta E=U_{I}^{(L)}=\langle \uparrow \mid \hat{H_{I}}(\frac{1}{E_{0}-\hat{H}%
_{0}}\hat{H_{I}})^{L_{0}-1}\left \vert \downarrow \right \rangle .
\end{equation}%
Due to the translation invariance, to calculate $(\frac{1}{E_{0}-\hat{H}_{0}}%
\hat{H_{I}})\left \vert \downarrow \right \rangle =(\frac{h^{x}}{E_{0}-\hat{H%
}_{0}}\sum \limits_{i}\sigma _{i}^{x})\left \vert \downarrow \right \rangle
, $ we can choose site $i$ as the starting point of the tunneling process
and get
\begin{eqnarray}
(\frac{1}{E_{0}-\hat{H}_{0}}\hat{H_{I}})\left \vert \downarrow \right
\rangle &\rightarrow &L_{x}L_{y}(\frac{h^{x}}{E_{0}-\hat{H}_{0}}\sigma
_{i}^{x})\left \vert \downarrow \right \rangle \\
&=&L_{x}L_{y}(\frac{h^{x}}{E_{0}-\hat{H}_{0}})\left \vert \Psi _{i}\right
\rangle  \nonumber
\end{eqnarray}%
where $\left \vert \Psi _{i}\right \rangle $ is the excited state of two $%
Z_{2} $ vortices (or $Z_{2}$ charges) at plaquettes $i-e_{y}$ and $i-e_{x}$
with an energy $E_{0}+4g$ (See Fig.4(a)). From $\hat{H}_{0}\left \vert \Psi
_{i}\right \rangle =(E_{0}+4g)\left \vert \Psi _{i}\right \rangle ,$ we have%
\[
(\frac{1}{E_{0}-\hat{H}_{0}}\hat{H_{I}})\left \vert \downarrow
\right
\rangle =L_{x}L_{y}(\frac{h^{x}}{-4g})\left \vert \Psi
_{i}\right
\rangle .
\]

In next step, one $Z_{2}$ vortex (or $Z_{2}$ charge) moves one step, we get

\begin{eqnarray}
&&(\frac{h^{x}}{E_{0}-\hat{H}_{0}}\sum \limits_{i}\sigma _{i}^{x})^{2}\left
\vert \downarrow \right \rangle  \nonumber \\
&=&(\frac{h^{x}}{E_{0}-\hat{H}_{0}}\sum \limits_{i}\sigma
_{i}^{x})L_{x}L_{y}(\frac{h^{x}}{-4g})\left \vert \Psi _{i}\right \rangle
\nonumber \\
&=&L_{x}L_{y}(\frac{h^{x}}{-4g})(\frac{h^{x}}{E_{0}-\hat{H}_{0}}\sum
\limits_{i}\sigma _{i}^{x})\left \vert \Psi _{i}\right \rangle  \nonumber \\
&=&L_{x}L_{y}(\frac{h^{x}}{-4g})(\frac{h^{x}}{E_{0}-\hat{H}_{0}}\sigma
_{i+e_{x}-e_{y}}^{x})\left \vert \Psi _{i}\right \rangle  \nonumber \\
&=&L_{x}L_{y}(\frac{h^{x}}{-4g})(\frac{h^{x}}{-4g})\left \vert \Psi
_{i}^{\prime }\right \rangle .
\end{eqnarray}%
where $\left \vert \Psi _{i}^{\prime }\right \rangle $ is the excited state
of two $Z_{2}$ vortices (or $Z_{2}$ charges) at plaquettes $i+e_{x}-2e_{y}$
and $i-e_{x}$. See Fig.4(b). Then step by step, one $Z_{2}$ vortex (or $%
Z_{2} $ charge) moves around the torus. When the $Z_{2}$ vortex (or $Z_{2}$
charge) goes back to its starting point and annihilates with the other, the
original quantum state $\left \vert \downarrow \right \rangle $ changes into
$\mid \uparrow \rangle $. Finally we get the energy splitting%
\begin{eqnarray}
\Delta &=&2\delta E=2U_{I}^{(L)}=2\langle \uparrow \mid \hat{H_{I}}(\frac{1}{%
E_{0}-\hat{H}_{0}}\hat{H_{I}})^{L_{0}-1}\left \vert \downarrow \right \rangle
\nonumber \\
&=&2\times L_{x}L_{y}\frac{(h^{x})^{L_{0}}}{(-4g)^{L_{0}-1}}=8L_{x}L_{y}g(%
\frac{h^{x}}{4g})^{L_{0}}.
\end{eqnarray}
It is noted that $L_{0}-1$ is an even number.

Because the quantum tunneling process of $Z_{2}$ vortex (or $Z_{2}$ charge)
plays a role of $\tau ^{x}$ on the quantum states $\left(
\begin{array}{c}
\mid \uparrow \rangle \\
\mid \downarrow \rangle%
\end{array}%
\right) $ as
\begin{equation}
\left(
\begin{array}{c}
\mid \downarrow \rangle \\
\mid \uparrow \rangle%
\end{array}%
\right) =\tau ^{x}\left(
\begin{array}{c}
\mid \uparrow \rangle \\
\mid \downarrow \rangle%
\end{array}%
\right) ,
\end{equation}%
we obtain the effective pseudo-spin Hamiltonian due to the contribution of $%
Z_{2}$ vortex (or $Z_{2}$ charges) as
\begin{equation}
\mathcal{\hat{H}}_{\mathrm{eff}}=\frac{\Delta }{2}(\mid \uparrow \rangle
\langle \downarrow \mid +\mid \downarrow \rangle \langle \uparrow \mid
)=J_{x}\tau ^{x}
\end{equation}%
where $J_{x}=\Delta /2$ \cite{kou1,kou1'}.

\begin{figure}[tbp]
\includegraphics[clip,width=0.5\textwidth]{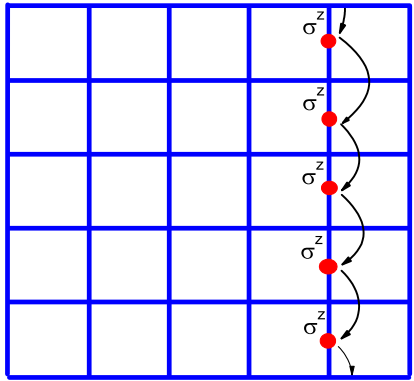}
\caption{Tunneling path of virtual fermion along $\hat{e}_{y}$-direction on
an $5\times 5$ lattice (The dots on the links denote the fermions). }
\label{Fig.5}
\end{figure}

For the second tunneling process, a virtual fermion will move around the
torus along direction $\hat{e}_{y}$ with length $L_{y}$ (It is noted that
due to $L_{x}\geq L_{y},$ the length of tunneling path along $\hat{e}_{x}$
direction is longer). See Fig.5. Such a tunneling process changes the
quantum states $\left(
\begin{array}{c}
\mid \uparrow \rangle \\
\mid \downarrow \rangle%
\end{array}%
\right) $ turn into $\left(
\begin{array}{c}
\mid \uparrow \rangle \\
-\mid \downarrow \rangle%
\end{array}%
\right) =\tau ^{z}\left(
\begin{array}{c}
\mid \uparrow \rangle \\
\mid \downarrow \rangle%
\end{array}%
\right) $. The extra sign of the state $\mid \downarrow \rangle $ comes from
the presence of $\pi $ flux of fermionic quasi-particles through the holes
of the torus. From Eq.\ref{sp}, we can get the energy shift of the state $%
\mid \downarrow \rangle $ as%
\begin{equation}
\delta E=\sum \limits_{j=0}^{\infty }\langle \downarrow \mid \hat{H}_{I}(%
\frac{1}{E_{0}-\hat{H}_{0}}\hat{H}_{I})^{j}\mid \downarrow \rangle
=L_{x}L_{y}\frac{(h^{z})^{L_{y}}}{(8g)^{L_{y}-1}}
\end{equation}
with an even number $L_{y}-1$. Through the same approach, we get the energy
shift $\Delta E$ of $\mid \downarrow \rangle $ is equal to $-L_{x}L_{y}\frac{%
(h^{z})^{L_{y}}}{(8g)^{L_{y}-1}}.$ Then an energy difference $\varepsilon $
of the two ground states is obtained as
\begin{equation}
\varepsilon =2\delta E=16L_{x}L_{y}g(\frac{h^{z}}{8g})^{L_{y}}.
\end{equation}%
\  \

Finally the two-level quantum system of the two degenerate ground states on
an $o\ast o$ lattice can be described by a simple effective pseudo-spin
Hamiltonian
\begin{eqnarray}
\mathcal{\hat{H}}_{\mathrm{eff}} &=&\frac{\Delta }{2}(\mid \uparrow \rangle
\langle \downarrow \mid +\mid \downarrow \rangle \langle \uparrow \mid )+%
\frac{\varepsilon }{2}(\mid \uparrow \rangle \langle \uparrow \mid -\mid
\downarrow \rangle \langle \downarrow \mid ) \\
&=&J_{x}\tau ^{x}+J_{z}\tau ^{z}  \nonumber
\end{eqnarray}%
where $J_{x}=\Delta /2$ and $J_{z}=\varepsilon /2$. By diagonalizing the
effective Hamiltonian matrix, we can get the eigenvalues of the two ground
states
\begin{equation}
E_{\pm }=\pm \sqrt{(\frac{\Delta }{2})^{2}+(\frac{\varepsilon }{2})^{2}}.
\end{equation}%
The total energy splitting becomes
\begin{equation}
\Delta E=E_{+}-E_{-}=2\sqrt{(\frac{\Delta }{2})^{2}+(\frac{\varepsilon }{2}%
)^{2}}.
\end{equation}%
For the Wen-plaquette model under external field along $x$-direction, the
total energy splitting $\Delta E$ is reduced into $\Delta =8L_{x}L_{y}g(%
\frac{h^{x}}{4g})^{L_{0}}.$ On the other hand, for the Wen-plaquette model
under external field along $z$-direction, the total energy splitting $\Delta
E$ is $\varepsilon =16L_{x}L_{y}g(\frac{h^{z}}{8g})^{L_{y}}.$

\subsection{Macroscopic quantum tunneling effect of the degenerate ground
states on $e\ast o$ lattice}

Secondly, we study the MQT of the two degenerate ground states on an $%
L_{x}\times L_{y}$ ($L_{x}$ is an even number and $L_{y}$ is an odd number)
lattice \cite{oe}. Now we map the two-fold degenerate ground states $\mid
m_{1}=0\rangle $ and $\mid m_{1}=1\rangle $ onto quantum states of the
pseudo-spin $\mathbf{\hat{\tau}}_{1}$ as $\mid \uparrow \rangle _{1}$ and $%
\left \vert \downarrow \right \rangle _{1}$, respectively. Under the
perturbation, $\hat{H}_{I}=h^{x}\sum \limits_{i}\sigma _{i}^{x}+h^{z}\sum
\limits_{i}\sigma _{i}^{z}$, there are two types of quantum tunneling
processes - virtual $Z_{2}$-vortex (or $Z_{2}$ charge) propagating along $%
\hat{e}_{x}-\hat{e}_{y}$ directions around the torus and virtual fermion
propagating along $\hat{e}_{y}$ direction around the torus.

For the virtual $Z_{2}$-vortex (or $Z_{2}$ charge) propagating along $\hat{e}%
_{x}-\hat{e}_{y}$ directions around the torus, the energy splitting $\Delta $
can be obtained by the high-order degenerate-state perturbation theory as
\begin{eqnarray}
\Delta &=&2\langle \uparrow \mid _{1}\hat{H_{I}}(\frac{1}{E_{0}-\hat{H}_{0}}%
\hat{H_{I}})^{L_{0}-1}\left \vert \downarrow \right \rangle _{1} \\
&=&2L_{x}L_{y}\frac{(h^{x})^{L_{0}}}{(-4g)^{L_{0}-1}}.  \nonumber
\end{eqnarray}
Because the quantum tunneling process of $Z_{2}$ vortex (or $Z_{2}$ charge)
plays a role of $\tau _{1}^{x}$ on the quantum states $\left(
\begin{array}{c}
\mid \uparrow \rangle _{1} \\
\mid \downarrow \rangle _{1}%
\end{array}%
\right) $ as $\left(
\begin{array}{c}
\mid \downarrow \rangle _{1} \\
\mid \uparrow \rangle _{1}%
\end{array}%
\right) =\tau _{1}^{x}\left(
\begin{array}{c}
\mid \uparrow \rangle _{1} \\
\mid \downarrow \rangle _{1}%
\end{array}%
\right) ,$ we obtain the effective pseudo-spin Hamiltonian due to the
contribution of $Z_{2}$ vortex (or $Z_{2}$ charge) as
\begin{equation}
\mathcal{\hat{H}}_{\mathrm{eff}}=\frac{\Delta }{2}(\mid \uparrow \rangle
_{1}\langle \downarrow \mid _{1}+\mid \downarrow \rangle _{1}\langle
\uparrow \mid _{1})=J_{x}\tau _{1}^{x}
\end{equation}%
where $J_{x}=\Delta /2$.

For the tunneling process of fermion propagating around the torus along
direction $\hat{e}_{y}$, we obtain the energy difference $\varepsilon $ of
the two ground states as
\begin{equation}
\varepsilon =2\Delta E=16L_{x}L_{y}g(\frac{h^{z}}{8g})^{L_{y}}.
\end{equation}%
The length of the tunneling path is $L_{y}$ which is an odd number. Such
tunneling process plays a role of $\tau _{1}^{z}$.
\begin{figure}[tbp]
\includegraphics[width=0.65\textwidth]{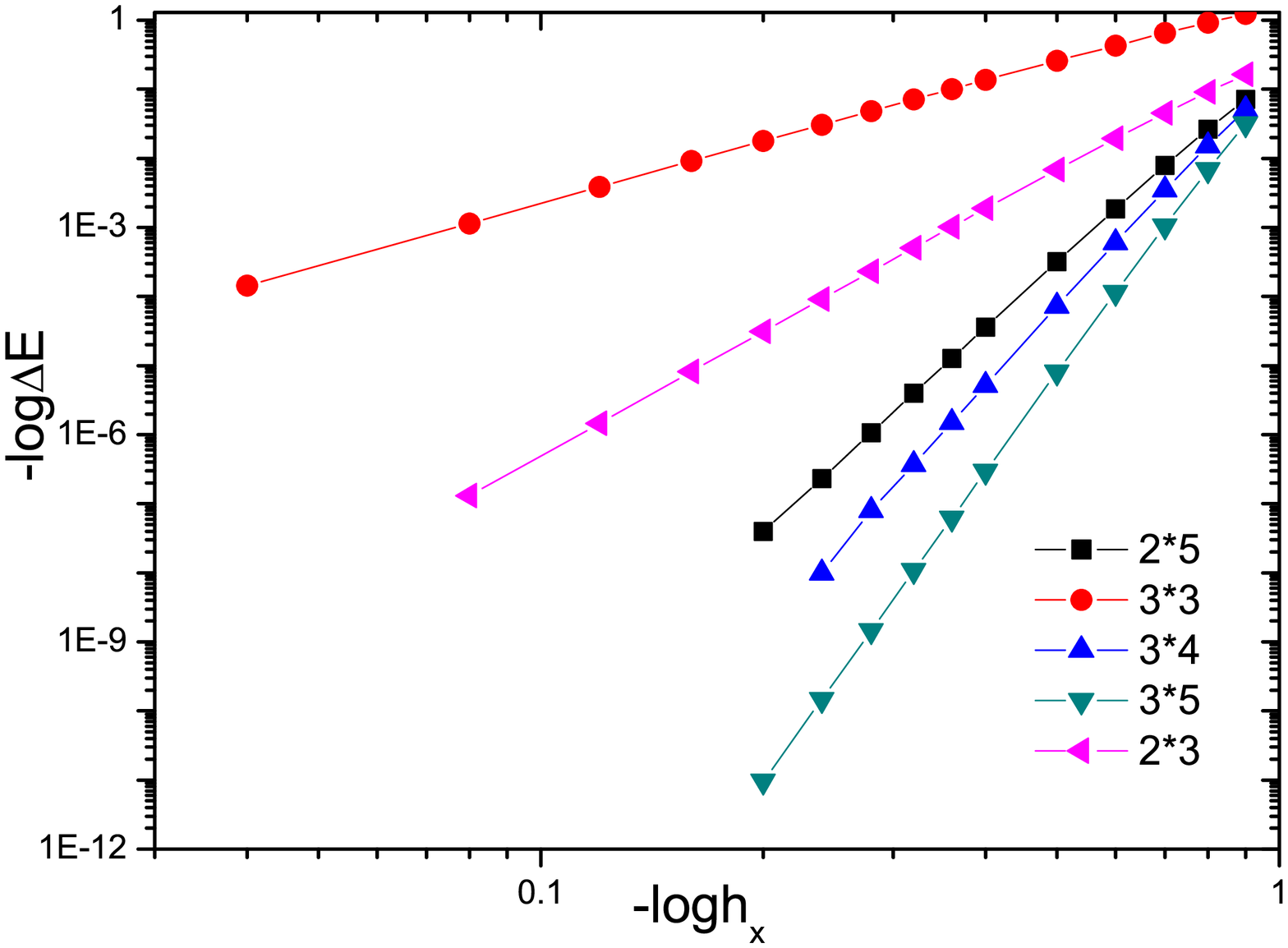}
\caption{The energy splitting between the two degenerate ground states of
the Wen-plaquette model in an external field along $x$-direction ($g=1$).
Here $N\ast M$ denotes a $N\times M$ lattice. }
\end{figure}

\begin{figure}[tbp]
\includegraphics[width=0.65\textwidth]{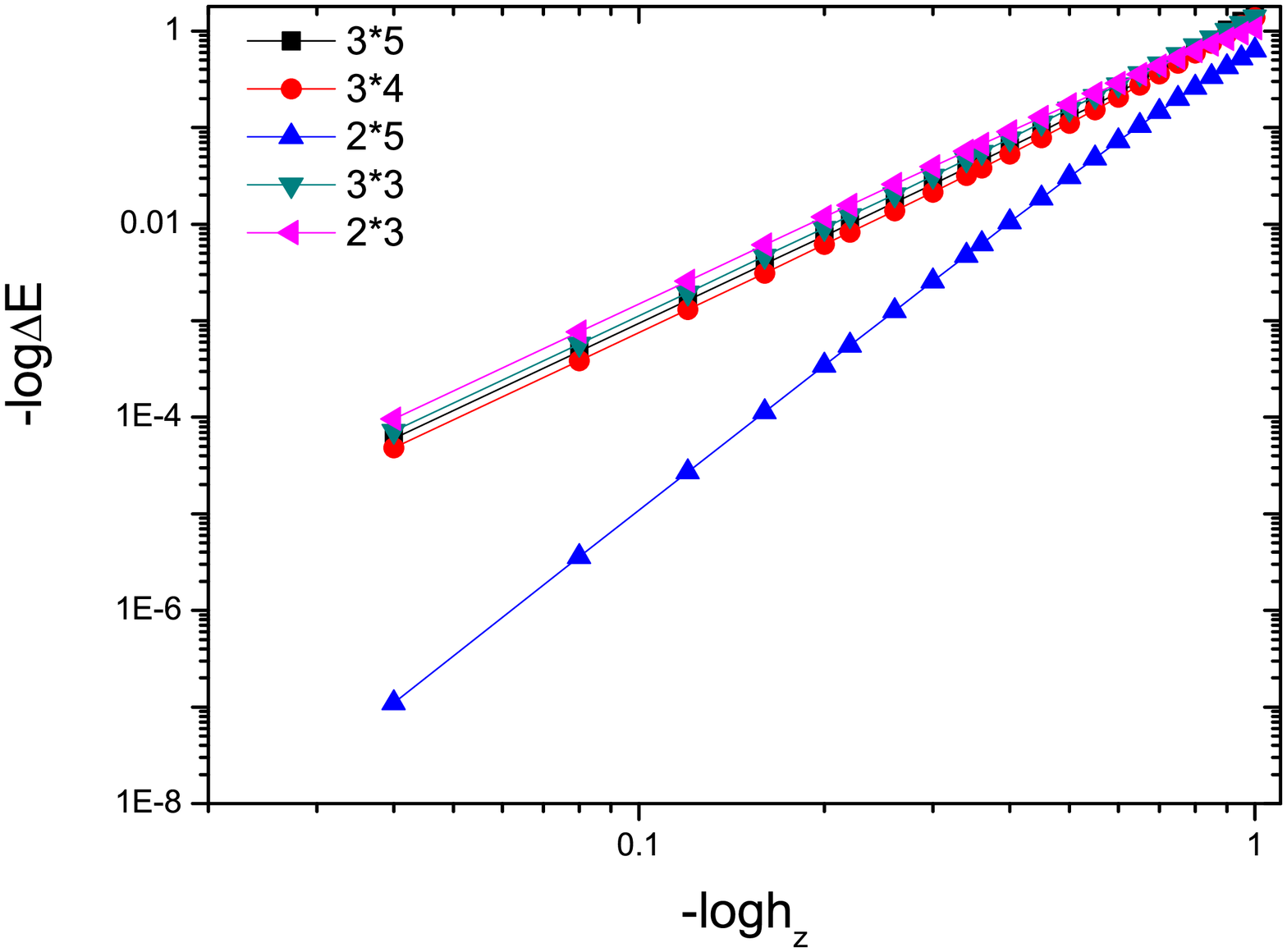}
\caption{The energy splitting between the two degenerate ground states of
the Wen-plaquette model in an external field along $z$-direction ($g=1$).
Here $N*M$ denotes a $N\times M$ lattice. }
\end{figure}

Finally the two-level quantum system of the two degenerate ground states on
an $e\ast o$ lattice can be described by
\begin{equation}
\mathcal{\hat{H}}_{\mathrm{eff}}=J_{x}\tau _{1}^{x}+J_{z}\tau _{1}^{z}
\end{equation}%
where $J_{x}=\Delta /2$ and $J_{z}=\varepsilon /2$. The total energy
splitting now becomes
\begin{equation}
\Delta E=E_{+}-E_{-}=2\sqrt{(\frac{\Delta }{2})^{2}+(\frac{\varepsilon }{2}%
)^{2}}.
\end{equation}%
In Fig.6 and Fig.7, we plot the numerical results from the exact
diagonalization technique of the Wen-plaquette model on different $o\ast o$
and $e\ast o$ lattices. Table.(IV) shows the tunneling lengths $L_{0}$ from
the numerical results (the numbers in the brackets are the theoretical
predictions), which indicate that our theoretical results are consistent
with the numerical results from exact diagonalization approach.

\begin{table}[t]
\begin{tabular}{|c|ccccc|}
\hline
$L_{0}$ & $3\ast 3$ & $2\ast 5$ & $3\ast 4$ & $3\ast 5$ &  \\ \hline
$h_{x}$ & 2.98312 (3) & 9.84653 (10) & 12.10754 (12) & 15.01707 (15) &  \\
\hline
$h_{z}$ & 3.06994 (3) & 4.83737 (5) & 3.03164 (3) & 3.01557 (3) &  \\ \hline
\end{tabular}%
\caption{The tunneling lengths $L_{0}$ from the numerical results (the
numbers in the brackets are the theoretical predictions). $h_{x}$ means the
external field along $x$-direction and $h_{z}$ means the external field
along $z$-direction. Here $N*M$ denotes a $N\times M$ lattice. }
\label{2}
\end{table}

\subsection{Macroscopic quantum tunneling effect of the degenerate ground
states on $e\ast e$ lattice}

Thirdly, we study the MQT of the four degenerate ground states on an $%
L_{x}\times L_{y}$ ($L_{x}$ and $L_{y}$ are even numbers with $L_{x}\geq
L_{y}$) lattice. We denote the four degenerate ground states $\mid
m_{1},m_{2}\rangle =$ $|0,0\rangle ,$ $|1,0\rangle ,$ $|0,1\rangle ,$ $%
|1,1\rangle $ by the quantum states of pseudo-spin $\mathbf{\hat{\tau}}_{1%
\text{ }}$and $\mathbf{\hat{\tau}}_{2}$. Under the perturbation, $\hat{H}%
_{I}=h^{x}\sum \limits_{i}\sigma _{i}^{x}+h^{z}\sum \limits_{i}\sigma
_{i}^{z}$, there are five types of quantum tunneling processes - virtual $%
Z_{2}$-vortex propagating along $\hat{e}_{x}-\hat{e}_{y}$ direction around
the torus, $Z_{2}$ charge propagating along $\hat{e}_{x}-\hat{e}_{y}$
direction around the torus, and virtual fermion propagating along $\hat{e}%
_{x},$ $\hat{e}_{y},$ $\hat{e}_{x}-\hat{e}_{y}$ direction around the torus,
respectively. We will calculate the ground state energy splitting\ from the
degenerate perturbation approach one by one.

In the first step we study the quantum tunneling process of $Z_{2}$-vortex
propagating along $\hat{e}_{x}-\hat{e}_{y}$ direction around the torus.
After such tunneling process, the quantum states $\left(
\begin{array}{c}
\mid 0,0\rangle \\
\mid 1,0\rangle \\
\mid 0,1\rangle \\
\mid 1,1\rangle%
\end{array}%
\right) $ turn into
\begin{equation}
\left(
\begin{array}{c}
\mid 0,0\rangle \\
\mid 1,0\rangle \\
\mid 0,1\rangle \\
\mid 1,1\rangle%
\end{array}%
\right) \rightarrow \left(
\begin{array}{c}
\mid 1,1\rangle \\
\mid 0,1\rangle \\
\mid 1,0\rangle \\
\mid 0,0\rangle%
\end{array}%
\right) =\tau _{1}^{x}\mathbf{\otimes }\tau _{2}^{x}\left(
\begin{array}{c}
\mid 0,0\rangle \\
\mid 1,0\rangle \\
\mid 0,1\rangle \\
\mid 1,1\rangle%
\end{array}%
\right) .
\end{equation}%
Thus we may use the pseudo-spin operator $\tau _{1}^{x}\mathbf{\otimes }\tau
_{2}^{x}$ to denote the tunneling process. The effective pseudo-spin
Hamiltonian due to the contribution of $Z_{2}$ vortex is obtained as
\begin{equation}
\mathcal{\hat{H}}_{\mathrm{eff}}=J_{xx}\tau _{1}^{x}\mathbf{\otimes }\tau
_{2}^{x}
\end{equation}%
where $J_{xx}=L_{x}L_{y}\frac{(h^{x})^{L_{0}}}{(-4g)^{L_{0}-1}}.$ Similar to
the results in above section, the length $L_{0}$ of the tunneling path is
equal to $\frac{L_{x}L_{y}}{\xi }$ where $\xi $ is the maximum common
divisor for $L_{x}$ and $L_{y}$.

In the second step we study the quantum tunneling process of $Z_{2}$-charge
propagating along $\hat{e}_{x}-\hat{e}_{y}$ directions around the torus. We
may use the pseudo-spin operator $\tau _{1}^{y}\mathbf{\otimes }\tau
_{2}^{y} $ to denote this tunneling process, of which the effective
pseudo-spin Hamiltonian is obtained as
\begin{equation}
\mathcal{\hat{H}}_{\mathrm{eff}}=J_{yy}\tau _{1}^{y}\mathbf{\otimes }\tau
_{2}^{y}
\end{equation}%
where $J_{yy}=L_{x}L_{y}\frac{(h^{x})^{L_{0}}}{(-4g)^{L_{0}-1}}$ and $L_{0}=%
\frac{L_{x}L_{y}}{\xi }.$

In the third step we study the quantum tunneling process of fermion
propagating along $\hat{e}_{x}$ and $\hat{e}_{y}$ directions around the
torus, of which the pseudo-spin operators correspond to $\mathbf{1}\otimes
\tau _{2}^{z}$ and $\tau _{1}^{z}\otimes \mathbf{1}$, respectively. Then the
effective pseudo-spin Hamiltonian due to the contribution of the two quantum
tunneling processes is obtained as
\begin{equation}
\mathcal{\hat{H}}_{\mathrm{eff}}=\tilde{h}_{1}^{z}\left( \tau _{1}^{z}%
\mathbf{\otimes 1}\right) +\tilde{h}_{2}^{z}\left( \mathbf{1\otimes }\tau
_{2}^{z}\right)
\end{equation}%
where $\tilde{h}_{1}^{z}=-8L_{x}L_{y}g(\frac{h^{z}}{8g})^{L_{x}}\ $and $%
\tilde{h}_{2}^{z}=-8L_{x}L_{y}g(\frac{h^{z}}{8g})^{L_{y}}$.

\begin{figure}[tbp]
\includegraphics[clip,width=0.65\textwidth]{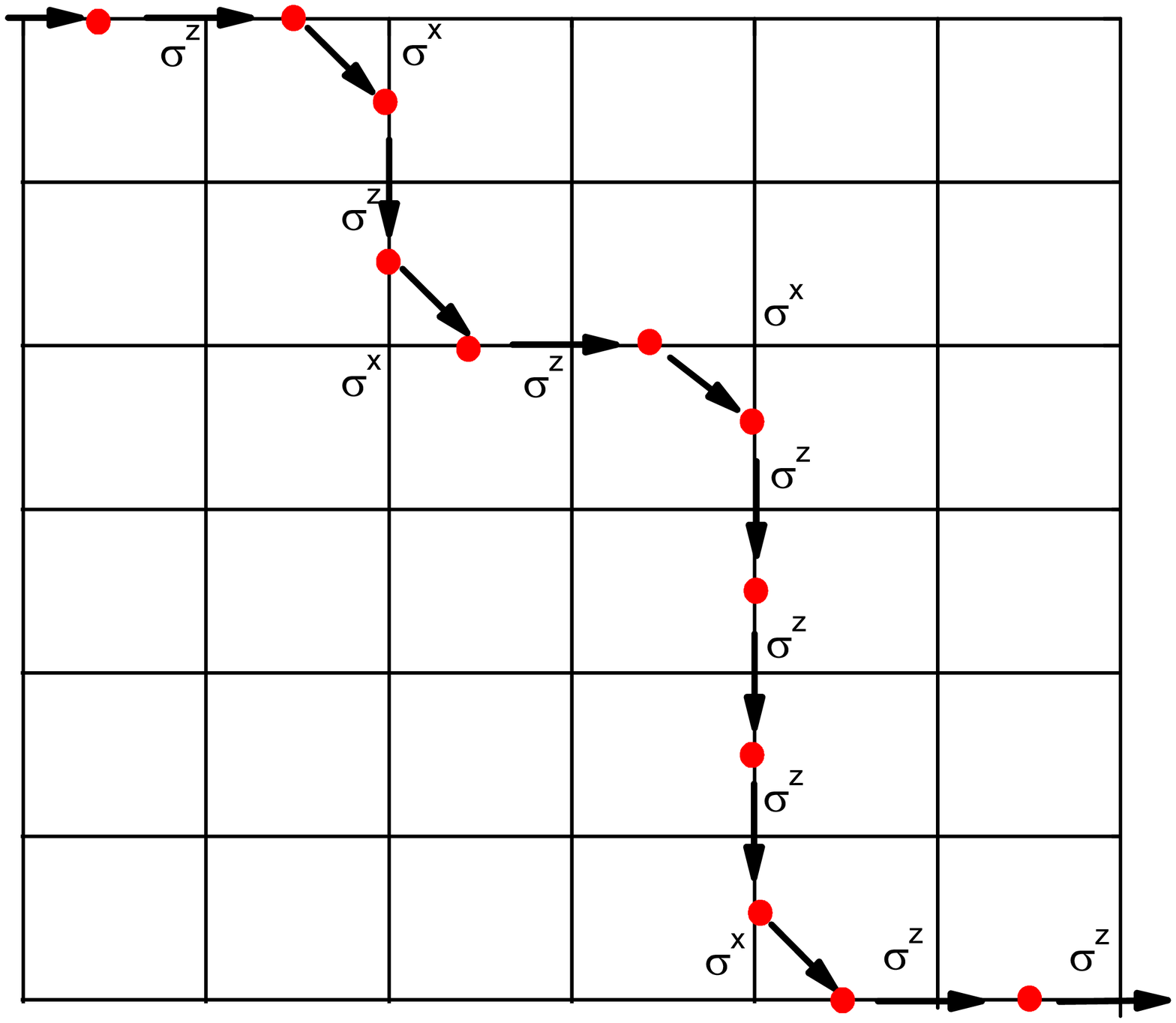}
\caption{Tunneling path of virtual fermions on an $6\times 6$ lattice with 4
corners (The dots on the links denotes fermions).}
\label{Fig.8}
\end{figure}

In the last step we study the quantum tunneling process of fermion
propagating along $\hat{e}_{x}-\hat{e}_{y}$ direction around the torus, of
which the pseudo-spin operator corresponds to $\tau _{1}^{z}\mathbf{\otimes }%
\tau _{2}^{z}$, respectively. Now there are a lot of tunneling pathes with
same length $2L_{0}$.\ Different tunneling pathes can be labeled by the
positions of corners, at which the fermions make a turn round from vertical
links to parallel links (or parallel links to vertical links). For a path
with $2k$\ corners ($k$ is an positive integer number), the topological
closed string operator can be written as%
\begin{equation}
W_{f}(\mathcal{C}_{XY})=\sigma _{i}^{z}\sigma _{i+1}^{z}...\sigma
_{j-1}^{z}\sigma _{j}^{x}\sigma _{j+1}^{z}...\sigma _{2L_{0}-2}^{x}\sigma
_{2L_{0}-1}^{z}\sigma _{2L_{0}}^{z}
\end{equation}%
with the site $i=(i_{x},i_{y})$ and a neighboring site $i+1$. See Fig.8.
Along the closed loops, each operator $\sigma _{j}^{x}$ corresponds to a
corner. Therefore, the number of pathes with $2k$\ corners that is equal to
the power of $h^{x}$ in $\Delta E$ is obtained as%
\begin{equation}
C_{L_{0}-1}^{k}=\frac{\left( L_{0}-1\right) !}{k!(L_{0}-k-1)!}.
\end{equation}%
\ It is noted that for any path, there are at least two corners. Then after
considering the tunneling processes of all possible pathes, the matrix
element of $\tau _{1}^{z}\mathbf{\otimes }\tau _{2}^{z}$ is obtained as
\begin{eqnarray}
J_{zz} &=&\varepsilon =L_{x}L_{y}C_{L_{0}-1}^{1}\frac{%
(2h^{x})^{2}(h^{z})^{2L_{0}-2}}{(-8g)^{2L_{0}-1}} \\
&&+L_{x}L_{y}C_{L_{0}-1}^{2}\frac{(2h^{x})^{4}(h^{z})^{2L_{0}-4}}{%
(-8g)^{2L_{0}-1}}+...  \nonumber \\
&&+L_{x}L_{y}C_{L_{0}-1}^{k}\frac{(2h^{x})^{2k}(h^{z})^{2L_{0}-2k}}{%
(-8g)^{2L_{0}-1}}  \nonumber \\
&&+...+L_{x}L_{y}\frac{(2h^{x})^{2L_{0}}}{(-8g)^{2L_{0}-1}}  \nonumber \\
&=&-L_{x}L_{y}\frac{[(2h^{x})^{2}+(h^{z})^{2}]^{L_{0}}-(h^{z})^{2L_{0}}}{%
(8g)^{2L_{0}-1}}.  \nonumber
\end{eqnarray}

Finally we derive an effective pseudo-spin Hamiltonian of the four ground
states as
\begin{widetext}
\begin{equation}
\mathcal{\hat{H}}_{\mathrm{eff}}\simeq J_{xx}\left( \tau _{1}^{x}\mathbf{%
\otimes }\tau _{2}^{x}\right) +J_{yy}\left( \tau _{1}^{y}\mathbf{\otimes }%
\tau _{2}^{y}\right) +J_{zz}\left( \tau _{1}^{z}\mathbf{\otimes }\tau
_{2}^{z}\right) +\tilde{h}_{1}^{z}\left( \tau _{1}^{z}\mathbf{\otimes 1}%
\right) +\tilde{h}_{2}^{z}\left( \mathbf{1\otimes }\tau _{2}^{z}\right)
\nonumber
\end{equation}
\end{widetext}which is equal to
\begin{widetext}
\begin{equation}
\mathcal{\hat{H}}_{\mathrm{eff}}=
  \left(\begin{array}{cccc}
     J_{zz}+\widetilde{h_{1}^{z}}+\widetilde{h_{2}^{z}} & 0 & 0 & J_{xx}-J_{yy} \\
     0 & -J_{zz}+\widetilde{h_{1}^{z}}-\widetilde{h_{2}^{z}} & J_{xx}+J_{yy} & 0 \\
    0 & J_{xx}+J_{yy} & -J_{zz}-\widetilde{h_{1}^{z}}+\widetilde{h_{2}^{z}} & 0 \\
     J_{xx}-J_{yy} &0 &0 &
     J_{zz}-\widetilde{h_{1}^{z}}-\widetilde{h_{2}^{z}}
  \end{array}\right).
\end{equation}\label{effm}
\end{widetext}The coefficients of $\mathcal{\hat{H}}_{\mathrm{eff}}$ are
given by
\begin{eqnarray}
J_{xx} &=&J_{yy}=L_{x}L_{y}\frac{(h^{x})^{L_{0}}}{(-4g)^{L_{0}-1}}, \\
J_{zz} &=&-L_{x}L_{y}\frac{%
[(2h^{x})^{2}+(h^{z})^{2}]^{L_{0}}-(h^{z})^{2L_{0}}}{(8g)^{2L_{0}-1}},
\nonumber \\
\tilde{h}_{1}^{z} &=&-8L_{x}L_{y}g(\frac{h^{z}}{8g})^{L_{x}},\text{ }\tilde{h%
}_{2}^{z}=-8L_{x}L_{y}g(\frac{h^{z}}{8g})^{L_{y}}.  \nonumber
\end{eqnarray}%
By diagonalizing the effective Hamiltonian, we get the energies of the
ground states as
\begin{eqnarray}
E_{1} &=&J_{zz}-\sqrt{(\tilde{h}_{1}^{z}-\tilde{h}_{2}^{z})^{2}+4J_{xx}^{2}},
\\
E_{2} &=&J_{zz}+\sqrt{(\tilde{h}_{1}^{z}-\tilde{h}_{2}^{z})^{2}+4J_{xx}^{2}},
\nonumber \\
E_{3} &=&J_{zz}+\tilde{h}_{1}^{z}+\tilde{h}_{2}^{z},  \nonumber \\
E_{4} &=&J_{zz}-\tilde{h}_{1}^{z}-\tilde{h}_{2}^{z}.  \nonumber
\end{eqnarray}%
Because the parameter $J_{zz}$ is always much smaller than others as $%
\left \vert J_{zz}\right \vert \ll \left \vert J_{xx}\right \vert ,$ $\left \vert
J_{yy}\right \vert ,$ $\left \vert \tilde{h}_{1}^{z}\right \vert ,$ $\left \vert
\tilde{h}_{2}^{z}\right \vert $, we may simplify $\mathcal{\hat{H}}_{\mathrm{%
eff}}$ as\cite{kou1}
\begin{equation}
\mathcal{\hat{H}}_{\mathrm{eff}}\simeq J_{xx}\left( \tau _{1}^{x}\mathbf{%
\otimes }\tau _{2}^{x}\right) +J_{yy}\left( \tau _{1}^{y}\mathbf{\otimes }%
\tau _{2}^{y}\right) +\tilde{h}_{1}^{z}\left( \tau _{1}^{z}\mathbf{\otimes 1}%
\right) +\tilde{h}_{2}^{z}\left( \mathbf{1\otimes }\tau _{2}^{z}\right)
\end{equation}%
and obtain the energies as
\begin{eqnarray}
E_{1} &\simeq &-\sqrt{(\tilde{h}_{1}^{z}-\tilde{h}_{2}^{z})^{2}+4J_{xx}^{2}},
\\
E_{2} &\simeq &\sqrt{(\tilde{h}_{1}^{z}-\tilde{h}_{2}^{z})^{2}+4J_{xx}^{2}},
\nonumber \\
E_{3} &\simeq &\tilde{h}_{1}^{z}+\tilde{h}_{2}^{z},  \nonumber \\
E_{4} &\simeq &-\tilde{h}_{1}^{z}-\tilde{h}_{2}^{z}.  \nonumber
\end{eqnarray}%
Then when the external field increases ($h^{x}\neq 0$ and $h^{z}\neq 0$),
the single energy level of the initial four degenerate ground states split
into four energy levels.

If we apply the external field along $z$-direction, the four energy levels
are%
\begin{eqnarray}
E_{1} &\simeq &-\tilde{h}_{1}^{z}+\tilde{h}_{2}^{z},\text{ }E_{2}\simeq
\tilde{h}_{1}^{z}-\tilde{h}_{2}^{z}, \\
E_{3} &\simeq &\tilde{h}_{1}^{z}+\tilde{h}_{2}^{z},\text{ }E_{4}\simeq -%
\tilde{h}_{1}^{z}-\tilde{h}_{2}^{z},  \nonumber
\end{eqnarray}%
where $\tilde{h}_{1}^{z}=-8L_{x}L_{y}g(\frac{h^{z}}{8g})^{L_{x}}$ and $%
\tilde{h}_{2}^{z}=-8L_{x}L_{y}g(\frac{h^{z}}{8g})^{L_{y}}.$\ In the
anisotropic limit, $L_{x}\gg L_{y}$, we have $\left \vert \tilde{h}%
_{1}^{z}\right \vert \ll \left \vert \tilde{h}_{2}^{z}\right \vert $. In this
case, the initial four degenerate ground states split into two groups, $%
E_{1}\simeq \tilde{h}_{2}^{z},$ $E_{2}\simeq -\tilde{h}_{2}^{z},$ $%
E_{3}\simeq \tilde{h}_{2}^{z}$ and $E_{4}\simeq -\tilde{h}_{2}^{z}.$ In each
group, there are two energy levels, of which the energy splitting $%
E_{1}-E_{3}=2\tilde{h}_{1}^{z}$ is very tiny. In contrast, the energy "gap"
between the two groups $E_{1}-E_{2}=-2\tilde{h}_{2}^{z}$ is larger. One can
see the energy levels of the Wen-plaquette model in external field along $z$%
-direction on $2\times 6$ lattice ($g=1$) in Fig.9. In the isotropic case, $%
L_{x}=L_{y}$, we have $\tilde{h}_{1}^{z}=\tilde{h}_{2}^{z}$. In this case,
the initial four degenerate ground states split into $E_{1}=E_{2}=0$, $%
E_{3}=2\tilde{h}_{1}^{z}$ and $E_{4}=-2\tilde{h}_{1}^{z}.$ One can see the
energy levels of the Wen-plaquette model in external field along $z$%
-direction on $4\times 4$ lattice ($g=1$) in Fig.10.
\begin{figure}[tbp]
\includegraphics[width=0.65\textwidth]{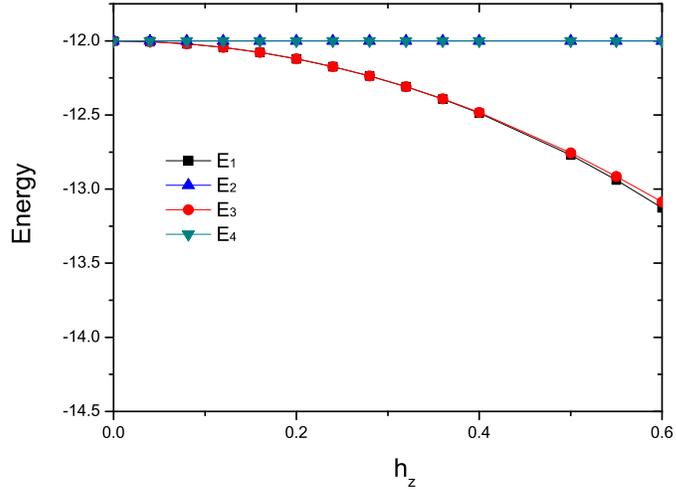}
\caption{The ground state energies of the Wen-plaquette model in an external
field along $z$-direction on $2\times 6$ lattice ($g=1$). }
\end{figure}

\begin{figure}[tbp]
\includegraphics[width=0.65\textwidth]{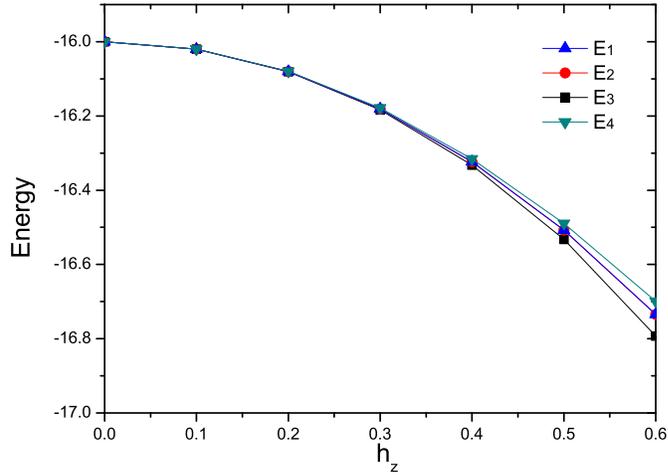}
\caption{The ground state energies of the Wen-plaquette model in an external
field along $z$-direction on $4\times 4$ lattice ($g=1$). }
\end{figure}

On the other hand, if we apply the external field along $x$-direction, the
four energy levels become%
\begin{eqnarray}
E_{1} &\simeq &-2J_{xx},\text{ }E_{2}\simeq 2J_{xx}, \\
E_{3} &=&E_{4}=J_{zz}\simeq 0,  \nonumber
\end{eqnarray}%
where $J_{xx}=L_{x}L_{y}\frac{(h^{x})^{L_{0}}}{(-4g)^{L_{0}-1}}$ and $%
J_{zz}=-8L_{x}L_{y}g(\frac{h^{x}}{8g})^{2L_{0}}$. Now the initial four
degenerate ground states split into three energy levels. One can see the
energy levels of the Wen-plaquette model in an external field along $x$%
-direction on $4\times 4$ lattice ($g=1$) in Fig.11.

\begin{figure}[tbp]
\includegraphics[width=0.65\textwidth]{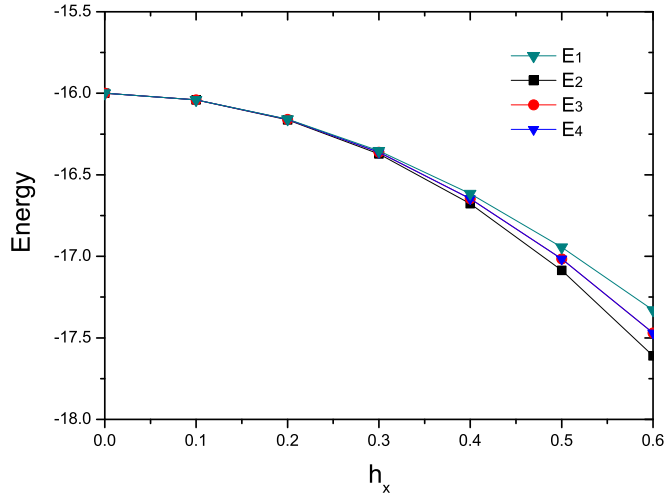}
\caption{The ground state energies of the Wen-plaquette model in an external
field along $x$-direction on $4\times 4$ lattice ($g=1$). }
\end{figure}

In addition, one may consider the MQT under a more general perturbation
\begin{equation}
\hat{H}_{I}=h^{x}\sum \limits_{i}\sigma _{i}^{x}+h^{y}\sum \limits_{i}\sigma
_{i}^{y}+h^{z}\sum \limits_{i}\sigma _{i}^{z}.  \label{xyz}
\end{equation}%
For an external field of $h^{x}\neq 0,$ $h^{y}\neq 0$ and $h^{z}\neq 0,$ all
quasi-particles ($Z_{2}$ vortex, $Z_{2}$ charge and fermion) can move along $%
\hat{e}_{x},$ $\hat{e}_{y},$ $\hat{e}_{x}\pm \hat{e}_{y}$ directions freely.
Therefore, to calculate the MQT of the degenerate ground states on an $e\ast
e$ lattice, all nine types of quantum tunneling processes should be
considered. The corresponding effective pseudo-spin Hamiltonian of the four
ground states turns into
\begin{eqnarray}
\mathcal{H}_{\mathrm{eff}} &=&J_{xx}\left( \tau _{1}^{x}\mathbf{\otimes }%
\tau _{2}^{x}\right) +J_{yy}\left( \tau _{1}^{y}\mathbf{\otimes }\tau
_{2}^{y}\right) +J_{zz}\left( \tau _{1}^{z}\mathbf{\otimes }\tau
_{2}^{z}\right)  \nonumber \\
&&+J_{zx}\left( \tau _{1}^{z}\mathbf{\otimes }\tau _{2}^{x}\right)
+J_{xz}\left( \tau _{1}^{x}\mathbf{\otimes }\tau _{2}^{z}\right) +\tilde{h}%
_{1}^{x}\left( \tau _{1}^{x}\mathbf{\otimes 1}\right)  \nonumber \\
&&+\tilde{h}_{2}^{x}\left( \mathbf{1\otimes }\tau _{2}^{x}\right) +\tilde{h}%
_{1}^{z}\left( \tau _{1}^{z}\mathbf{\otimes 1}\right) +\tilde{h}%
_{2}^{z}\left( \mathbf{1\otimes }\tau _{2}^{z}\right)
\end{eqnarray}%
where $J_{xx},$ $J_{yy}$, $J_{zz},$ $J_{zx}$, $J_{xz},$ $\tilde{h}_{1}^{x},$
$\tilde{h}_{2}^{x}$, $\tilde{h}_{1}^{z},$ $\tilde{h}_{2}^{z}$ are determined
by the energy splitting of the degenerate ground states from the nine
tunneling processes \cite{kou1,kou1'}. This issue (the MQT of Eq. (\ref{xyz}%
)) will be studied elsewhere.

\section{Conclusion}

In this paper, we study macroscopic quantum tunneling (MQT) effect of $Z_{2}$
topological order in the Wen-Plaquette model that is characterized by the
quantum tunneling processes of different virtual quasi-particles moving
around the torus. By focusing on the degenerate ground states, we get their
effective pseudo-spin models. The coefficients of these effective
pseudo-spin models are obtained by a high-order degenerate perturbation
approach. With the help of the effective pseudo-spin models, the energies of
the ground states are calculated and the results are consistent with those
from exact diagnalization numerical technique.

In the future, the approach will be applied onto the MQTs of $Z_{2}$
topological order in other models, such as the Kitaev toric-code mode and
the Kitaev model on honeycomb lattice. After learning the nature of the MQT
of $Z_{2}$ topological orders in different models, one may know how to
manipulate the degenerate ground states by controlling the external field
and then do topological quantum computation within the degenerate ground
states \cite{kou1,kou1'}.

\begin{acknowledgments}
This research is supported by NCET, NFSC Grant no. 10874017.
\end{acknowledgments}


\begin{thebibliography}{99}
\bibitem{oe} The MQT on $o\ast e$ lattice is identical to that on $e\ast o$
lattice, except that the two degenerate states are denoted by $|0,0\rangle ,$
$|0,1\rangle \ $rather than $|0,0\rangle ,$ $|1,0\rangle $.
\end{thebibliography}
\end{document}